\documentclass[aps,prb,twocolumn,superscriptaddress]{revtex4-2}
\usepackage{graphicx}
\usepackage{bm}
\usepackage{amsmath}
\usepackage{amssymb}
\usepackage{amsfonts}
\usepackage{euscript}
\usepackage{verbatim}
\usepackage{setspace}
\usepackage{xcolor}
\usepackage{amsfonts}
\usepackage{braket}

\newcommand{\be}{\begin{equation}}
\newcommand{\ee}{\end{equation}}
\newcommand{\bea}{\begin{eqnarray}}
\newcommand{\eea}{\end{eqnarray}}

\newcommand{\Renyi}[0]{R\'{e}nyi~}

\newcommand{\Tr}[1]{\mathrm{Tr} #1}

\newcommand{\calh}{\mathcal{H}}
\newcommand{\intz}{\mathbb{Z}}

\begin{document}

\title{Symmetry-resolved entanglement in symmetry-protected topological phases}
\author{Daniel Azses}
\affiliation{Department of Physics, Bar-Ilan University, Ramat Gan 5290002, Israel}
\affiliation{Center for Quantum Entanglement Science and Technology, Bar-Ilan University, Ramat Gan 5290002, Israel}

\author{Eran Sela}
\affiliation{School of Physics and Astronomy, Tel Aviv University, Tel Aviv 6997801, Israel}

\begin{abstract}
Symmetry protected topological phases (SPTs) have universal degeneracies in the entanglement spectrum in one dimension (1D). Here, we formulate this phenomenon in the framework of symmetry-resolved entanglement (SRE) using cohomology theory. We develop a general approach to compute  entanglement measures of SPTs in any dimension and specifically SRE via a discrete path integral on multi-sheet Riemann surfaces with generalized defects. The resulting path integral is expressed  in terms of group cocycles describing the topological actions of SPTs. Their cohomology classification allows to identify universal entanglement properties. Specifically, we demonstrate an equi-block decomposition of the reduced density matrix into symmetry sectors, for all 1D topological phases protected by finite Abelian unitary symmetries.
\end{abstract}

\maketitle

\section{Introduction}
Symmetry protected topological  phases (SPTs) are quantum mechanical states of matter respecting a symmetry and having a finite energy gap. Nontrivial SPTs have fractionalized edge states~\cite{affleck2004rigorous} and a peculiar form of short range entanglement, making them resource states for measurement-based quantum computation~\cite{else2012symmetry,stephen2017computational}. 
Specifically, the appearance of topologically protected degeneracies in the entanglement spectrum in one-dimension~\cite{pollmann2010entanglement} (1D) is a key property which is invariant under adiabatic deformations of the wave-function~\cite{chen2010local,fidkowski2011topological,chen2013symmetry}. 

In this work we study SPTs in the framework of symmetry-resolved entanglement (SRE)~\cite{goldstein2018symmetry,laflorencie2014spin,xavier2018equipartition,tan2019particle,feldman2019dynamics,bonsignori2019symmetry,fraenkel2020symmetry,horvath2020symmetry,calabrese2020full,murciano2020symmetry,murciano2020entanglement,turkeshi2020entanglement}. Consider a ground state $|\Psi \rangle$ of a Hamiltonian respecting a conservation law leading to conserved charge, denoted $\mathcal{Q}$, e.g. the total number of particles in the system. The full system has a fixed total charge, but for a bipartition of the system into two regions $A$ and $B$, the charge of each region may fluctuate. Yet,  the reduced density matrix $\rho_A={\rm{Tr}}_B |\Psi \rangle \langle 
\Psi |$, and hence its spectrum $\lambda_i$, i.e. the entanglement spectrum, can be block-decomposed into symmetry sectors associated with the conserved charge $\mathcal{Q}$ in the subregion $A$. This allows to symmetry-resolve the entanglement entropy $S=-\sum_i \lambda_i \log \lambda_i$ or its various moments $s_n =\sum_i \lambda_i^n$ (``\Renyi entropy"). The entanglement spectrum stemming from symmetry sector $\mathcal{Q}$  is obtained by applying a projector operator $P_\mathcal{Q}$ to a given charge $\mathcal{Q}$ of  subsystem $A$, $\{ \lambda_i \}_\mathcal{Q} =  {\rm{spec}} \rho_A P_\mathcal{Q}$. SRE was addressed in a number of topological systems hosting non-Abelian anyons~\cite{cornfeld2019entanglement}, in SPTs~\cite{de2020inaccessible}, and it can also  be measured experimentally~\cite{cornfeld2018imbalance,cornfeld2019measuring} as demonstrated recently on an IBM quantum computer~\cite{azses2020identification}.

Here we are interested in the decomposition of entanglement of general SPTs according to the underlying protecting symmetry. SPT ground states  are invariant under the  action of a symmetry
\be
\label{eq:localsymmetry}
u(g) \otimes \dots \otimes u(g) | \Psi \rangle= | \Psi \rangle,
\ee 
where the product is over sites on a lattice, $g \in G$ is an element of the symmetry group $G$ protecting the SPT, and $u(g)$ is an on-site representation of the symmetry. %
For unitary symmetries Eq.~(\ref{eq:localsymmetry}) is associated with a conserved charge. One can  project  into the   generalized charge sectors which, for Abelian finite groups, can be written in terms of the group characters $\chi_{\mathcal{Q}}(g)$ \cite{yen2019exact, de2020inaccessible},
\be
\label{charac}
P_\mathcal{Q}=\frac{1}{|G|} \sum_{g \in G} {\chi_\mathcal{Q}(g)} U_A(g).
\ee 
Here $U_A(g)= \otimes_{i \in A} u(g)_i$ acts only on subsystem $A$.  For the finite Abelian groups we shall consider, charge sectors $\mathcal{Q}$ are group elements $\mathcal{Q} \in G$.

As a new tool to extract universal information about the entanglement SPTs, in this paper we develop a discrete path integral approach to compute the SRE of SPTs. We  build on topological actions which are believed to provide a full description of SPTs in terms of group cocycles and their cohomological classification~\cite{chen2013symmetry}. %
While the method allows to extract entanglement properties of SPTs in any dimension and symmetry, here we 
concentrate on Abelian finite groups in 1D.

Focusing on 1D SPTs, we find that nontrivial SPTs   generically display equi-block decomposition, meaning that entanglement spectra of different symmetry blocks $\{ \lambda_i \}_\mathcal{Q}$ are degenerate. We note that a weaker notion of entanglement equipartition was put forward by Xavier {\emph{et al.}}~\cite{xavier2018equipartition}, see also Refs.~\cite{horvath2020symmetry,calabrese2020full,turkeshi2020entanglement}, in the context of conformal as well as gapped field theories. There, symmetry blocks of the reduced density matrix are first normalized to be legitimate density matrices $\frac{\rho_A P_\mathcal{Q}}{{\rm{tr}} \rho_A P_\mathcal{Q}}$, which then turn out to have an identical low energy structure. Our stronger notion of equidecomposition is reflected by the decomposition into identical blocks, {\emph{i.e.}}, in SPTs the probability to be found in various symmetry sectors are identical.

For certain SPT  phases, the equidecomposition is complete, i.e. the spectra $\{ \lambda_i \}_\mathcal{Q}$ is independent of $\mathcal{Q}$. We also identify topological phases with a partial degeneracy between symmetry sectors. For example this occurs for symmetry groups $G=\intz_N \times \intz_N$ where $N$ is not prime. The entanglement equidecomposition provides a relationship between SRE and  the degeneracies in the entanglement spectrum~\cite{pollmann2010entanglement}.
The method allows to study nonuniversal features that vary withing topological phases, by studying co-boundary transformations. While the entanglement entropy itself in nonuniversal, it has minimal value which is a property of each SPT~\cite{de2020inaccessible}, and is intimately connected with SRE equidecomposition.

\begin{figure*}[t]
	\centering
	\includegraphics[scale=0.4]{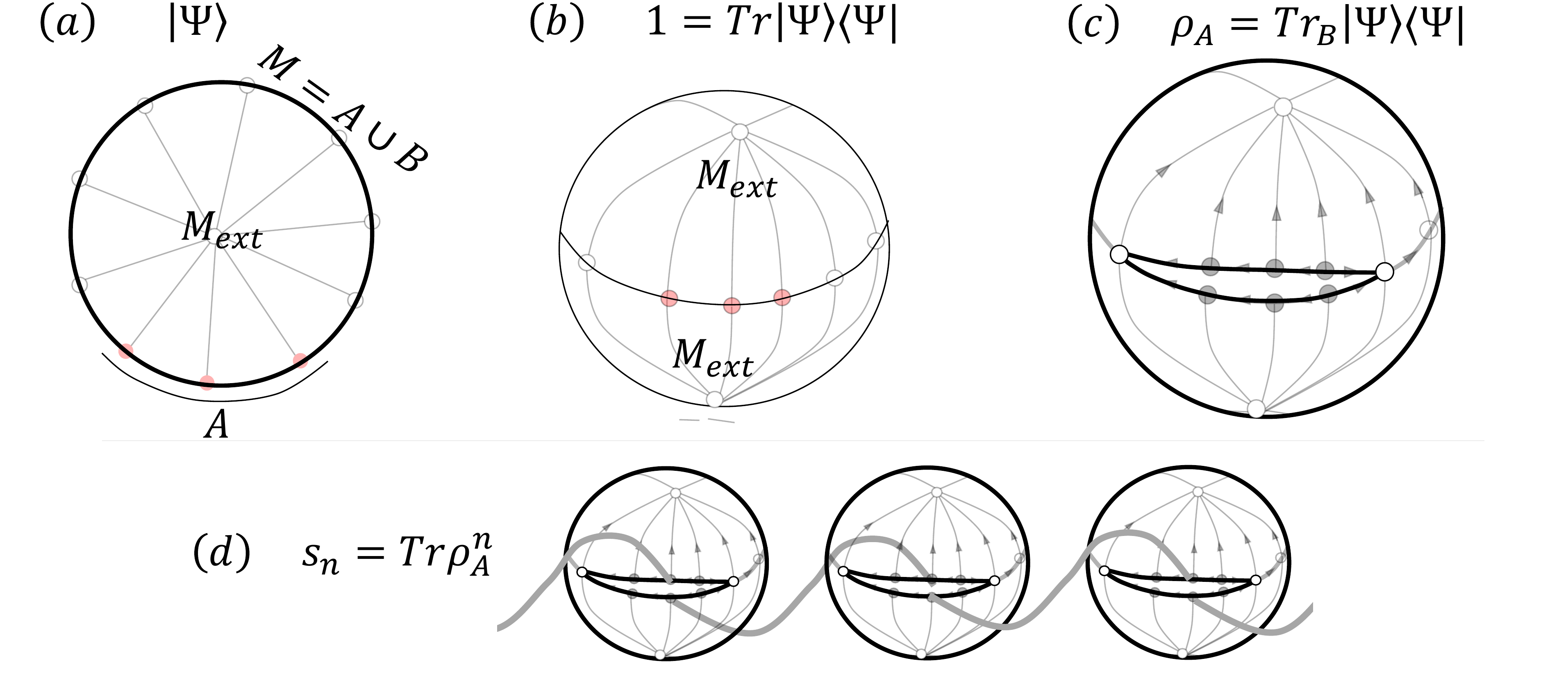}
	\caption{(a) The wave function of a $d=1$ SPT on a circle $(M=S^d)$ as an action amplitude on an extended manifold $M_{ext}$ (a $d+1-$ball) with $M=\partial M_{ext}$. (b) The normalization condition ${\rm{Tr}} |\Psi \rangle \langle \Psi |=1$ is expressed as a trivial action amplitude on a closed manifold of a $d+1$-sphere obtained by gluing a pair of $M_{ext}$ manifolds at their boundary. (c) The reduced density matrix corresponds to an action amplitude on an open manifold, whose boundary is the union of the $A$ subsystems from each $M_{ext}$. (d) The $n$-th \Renyi entropy as an action amplitude on an $n$-sheet Riemann surface.}
	\label{fig1}
\end{figure*}
\section{SRE of SPTs from topological path integral}
Entanglement measures can be represented  using quantum field theory as a path integral on multi-sheet Riemann surfaces~\cite{holzhey1994geometric,calabrese2004entanglement}. SRE can be incorporated by introducing generalized Aharonov-Bohm fluxes into this space~\cite{goldstein2018symmetry,xavier2018equipartition}. For theories like 1D conformal field theories the resulting partition functions can be computed exactly~\cite{goldstein2018symmetry,xavier2018equipartition}. Progress was also made using gapped theories~\cite{murciano2020symmetry,murciano2020entanglement}. Here we deal with gapped theories comprising of topological Wess-Zumino-Witten (WZW)-like terms~\cite{wess1971consequences,witten1994non} which were argued to give a general description of SPT phases~\cite{chen2013symmetry}. 

The key feature of the employed field theories 
representing SPTs, that will be specified in a discrete form in Sec.~\ref{se:triang}, is, \emph{property (i)}, that they always give a trivial action amplitude 
\be
\label{property1}
e^{-\int_{{\rm{closed}}} d^{d+1} x   \mathcal{L}[g(x)]}=1
\ee
 for a closed manifold~\cite{genus}. Chen \emph{et. al.}~\cite{chen2013symmetry} formulated   general fixed point wave functions of a $d-$dimensional SPT living on a closed manifold $M$, by arbitrarily extending $M$ to be the boundary of a $d+1$-dimensional manifold, $M=\partial M_{ext}$,
 \be
 \label{eq:WENwf}
 \psi (g(x))|_{x \in M}=\int_{M_{ext}} Dg e^{-\int d^{d+1} x  \mathcal{L}[g(x)]},
 \ee
 with the boundary condition that the field coincides with $g(x)$ on $M$. 
 Assuming periodic boundary conditions (PBC), we take $M$ and $M_{ext}$ to be a $d-$sphere and a $d+1$-ball, respectively, as depicted in Fig.~\ref{fig1}(a) for $d=1$.
Since the extension of $M$ into $M_{ext}$ is arbitrary, the theory also satisfies  \emph{property (ii)}: the action amplitude depends only on the field on the boundary. %
 Together with the symmetry condition 
 \be
 \label{eq:sym}
 e^{-\int d^{d+1} x  \mathcal{L}[g(x)]}=e^{-\int d^{d+1} x  \mathcal{L}[gg(x)]}, ~~~g \in G,
 \ee Chen \emph{et. al.}~\cite{chen2013symmetry} argued that the classification of these field theories is equivalent to that of SPTs. %

The normalization condition $\langle \Psi|\Psi \rangle =1$ or ${\rm{Tr}}| \Psi \rangle \langle \Psi |=1$ is then trivially represented from Eq.~(\ref{property1}) by the path integral over the closed surface obtained by gluing a pair of $d+1$-balls 
on their boundaries, resulting in a closed manifold equivalent to a $d+1$-sphere, see Fig.~\ref{fig1}(b). If we divide $M$ into regions $A$ and $B$, which we take to be equivalent to the two halves of the $d-$sphere, the reduced density matrix  ${\rm{Tr}}_B| \Psi \rangle \langle \Psi |$ is represented as path integral on a manifold with a boundary, as depicted in Fig.~\ref{fig1}(c).  The $n-$th \Renyi entropy is then represented as a path integral over the $n-$sheet Riemann surface, see Fig.~\ref{fig1}(d).

\begin{figure}[t]
	\centering
	\includegraphics[scale=0.4]{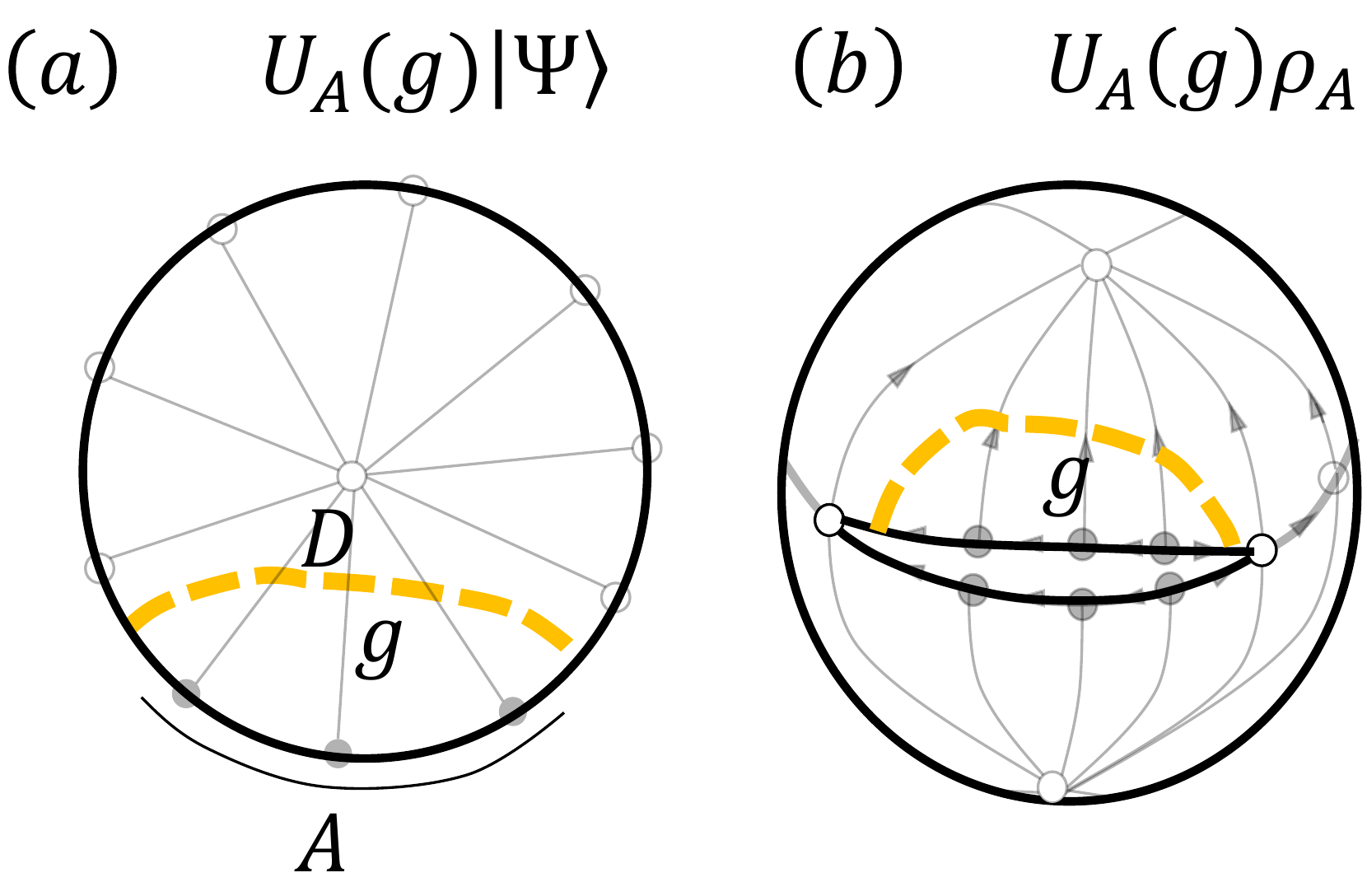}
	\caption{(a) A defect line $\mathcal{D}$ is attached at a pair of points $\partial A=\partial D$ to $M$ and extends arbitrarily through $M_{ext}$. We implement the transformation $g(x) \to g'(x)=g g(x)$ for $x \in M_{ext}$ located on one side (marked $g$) of the defect. The resulting action amplitude $e^{-\int d^{d+1} x  \mathcal{L}[g'(x)]}$ represents $U(g)|\Psi \rangle$. (b) Corresponding manifold and added defect for $U_A(g) \rho_A$. }
	\label{fig2}
\end{figure}

To obtain the SRE we apply the projector onto a given symmetry sector. To do so we  assume that projectors can be written as in Eq.~(\ref{charac}) in terms of the symmetry operators $U_A(g)= \otimes_{i \in A} u(g)_i$. This requires to apply the transformation $U_A(g)$ on the wave function. In the $|g \rangle$  basis this amounts to taking $g(x) \to g g(x)$ for $x \in A$. This can be readily implemented in the action amplitude expression Eq.~(\ref{eq:WENwf}) in the extended manifold. Note that the action amplitude is invariant under a global symmetry transformation Eq.~(\ref{eq:sym}), and also, due to property (ii), it is also invariant under  any local transformation  of the field configuration $g(x)\to g'(x)=g g(x)$ acting only inside $M_{ext}$, i.e. not on the boundary, $M$. On the contrary, consider a $d-$dimensional defect $\mathcal{D}$, living in $M_{ext}$, whose boundary $\partial \mathcal{D}$ is in $M$ and which coincides with the boundary of $A$, $\partial \mathcal{D}=\partial \mathcal{A}$. This is illustrated in Fig.~\ref{fig2}(a) for $d=1$, in which case $\mathcal{D}$ is a line defect extending through $M_{ext}$ from the pair of end points of $A$. Applying the transformation on a submanifold of $M_{ext}$ bounded by  $\mathcal{D}$ and $A$, gives the %
wave function 
\be
(U(g) \psi)(g(x))=\int_{M_{ext}} Dg e^{-\int d^{d+1} x  \mathcal{L}[g'(x)]},
\ee
$g'(x)= g^{-1} g(x)|_{x~{\rm{bounded~by}}~\mathcal{D},A}$.
Similarly, we can construct a path integral expression for the symmetry reduced density matrix, Fig.~\ref{fig2}(b), where symmetry resolution requires to use characters as in Eq.~(\ref{charac}).

\subsection{Discrete space: complexes and cocycles}
\label{se:triang}
Chen \emph{et. al.}~\cite{chen2013symmetry} proposed a field theory due to Dijkgraaf and Witten~\cite{dijkgraaf1990topological} satisfying Eqs.~(\ref{property1}),~(\ref{eq:sym}) and the associated properties (i) and (ii), written in a discrete $d$-dimensional triangulated space - a complex. See also Ref.~\cite{mesaros2013classification} for clarifying  discussions; %
Properties (i) and (ii) are encoded there~\cite{mesaros2013classification} as theorems 1 and 2.   

Consider a triangulation of the manifold $M$ into elementary $d+1$-dimensional simplexes, see Figs.~1,2. Attaching a ``spin" variable $g_i \in G$ to each vertex, our action amplitude is  a product over all the elementary $d+1$-simplexes over the $U(1)$-valued function%
\be
e^{-\int d^{d+1} x   \mathcal{L}[g(x)]} \to \prod_{ij\dots k}\nu_{1+d}^{s_{ij \dots k}}(g_i,g_j,\dots,g_k).
\ee
The key object here is the group cocycle $\nu_{1+d}(g_0,g_1,\dots,g_{d+1})$ being a $U(1)$-valued function of $d+2$ variables, that satisfy (a) the symmetry condition 
\be
\label{eq:symcocycle}
\nu_{1+d}(g_0,g_1,\dots,g_{d+1})=\nu_{1+d}(g g_0,g g_1,\dots,g g_{d+1}),
\ee 
and (b) that a product of cocycles over any closed $d+1$ manifold is trivial~\cite{genus}
\be
\label{cocycle}
\prod_{ij\dots k}\nu_{1+d}^{s_{ij \dots k}}(g_i,g_j,\dots,g_k)|_{closed~manifold}=1.
\ee  The latter is called the cocycle condition, equivalent to Eq.~(\ref{property1}). The complex has a branching structure that determines the values of $s_{ij \dots k}=\pm 1$~\cite{chen2013symmetry}. Having found a cocycle in $d+1$-dimensions satisfying Eqs.~(\ref{eq:symcocycle}) and (\ref{cocycle}), one can perform a co-boundary transformation, simply by attaching to each  $d$-dimensional simplex at the boundary of each $d+1$-dimensional simplex an arbitrary function $\mu_d(g_0,\dots,g_d)$ that satisfies the symmetry condition Eq.~(\ref{eq:symcocycle}). This results in an equivalent cocycle. So, co-boundary transformations define equivalence classes of cocycles. The cohomology group $\mathcal{H}^{1+d}(G,U(1))$ classifies cocycles up to co-boundary transformations. The fundamental conjecture of Chen \emph{et. al.}~\cite{chen2013symmetry} is that this classifies SPTs into phases, also yielding explicit form for their wave functions~\cite{chen2013symmetry}. %

\section{Entanglement spectrum in 1D from cohomology}
The field theory satisfying property (ii) on  complexes leads to \emph{triangulation invariance}: the action amplitude does not depend  on the internal triangulation of the complex~\cite{chen2013symmetry}. This allows us to express universal (and nonuniversal) entanglement measures in terms of a minimal number of cocycles involving the edge of the $n$-sheet Riemann surfaces introduced above. Nonuniversal properties are those that depend on co-boundary transformations. To demonstrate these ideas we now turn to 1D SPTs.

Focusing on systems with PBC, the reduced density matrix in Fig.~\ref{fig1}(c) for a chain with $L_A$ sites is equivalent to a 2-ball (a disk),
\bea
\label{eq:sn1}
\raisebox{-.5\height}{\includegraphics[scale=0.3]{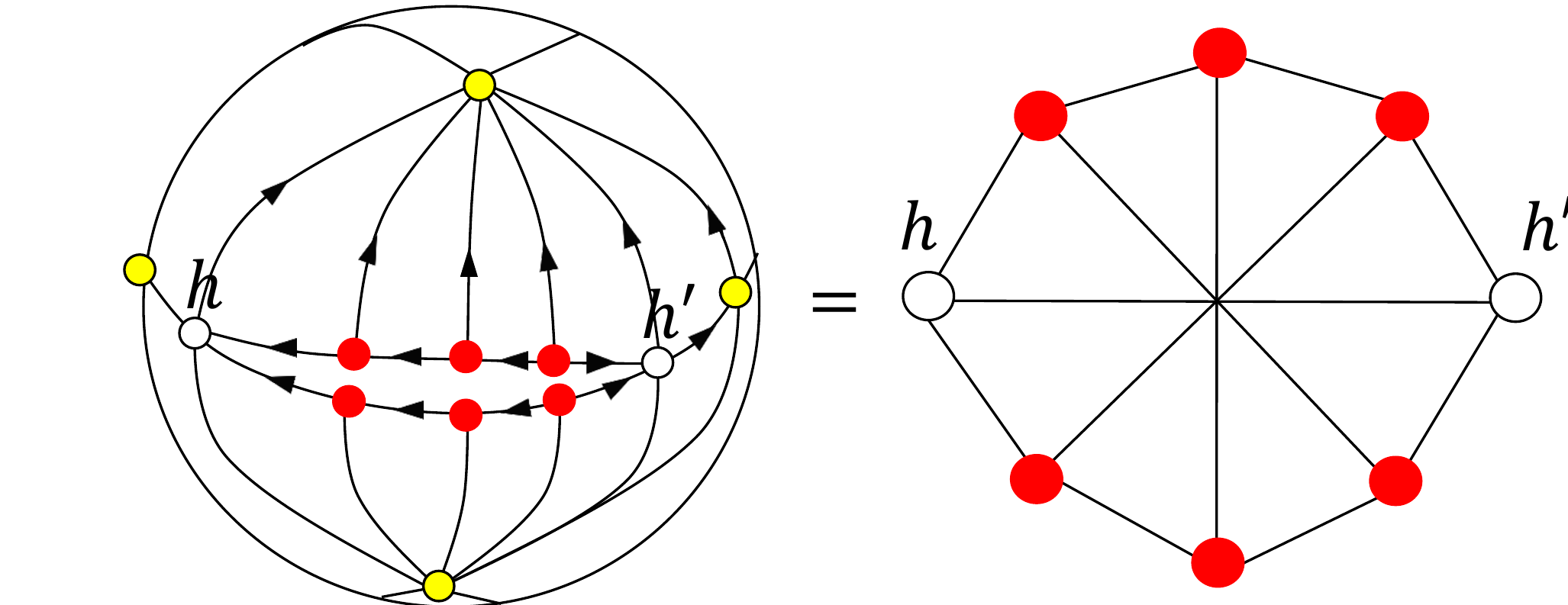}}.\nonumber
\eea
We use $\{g\}$ to denote sites in $A$ ($2L_A$ sites in total for $\rho_A$) and $\{h \}$ for the rest. The latter include two sites $h,h'$ on the boundary of the manifold originating from subsystem $B$, as well as internal vertices. %

The $n-$th \Renyi entropy  in Fig.~\ref{fig1}(c)  is obtained by  identifying the upper edge of the $i-$th disk with the lower edge of the $i+1$-th disk periodically,
\bea
\label{eq:sn2}
&&\raisebox{-.5\height}{\includegraphics[scale=0.4]{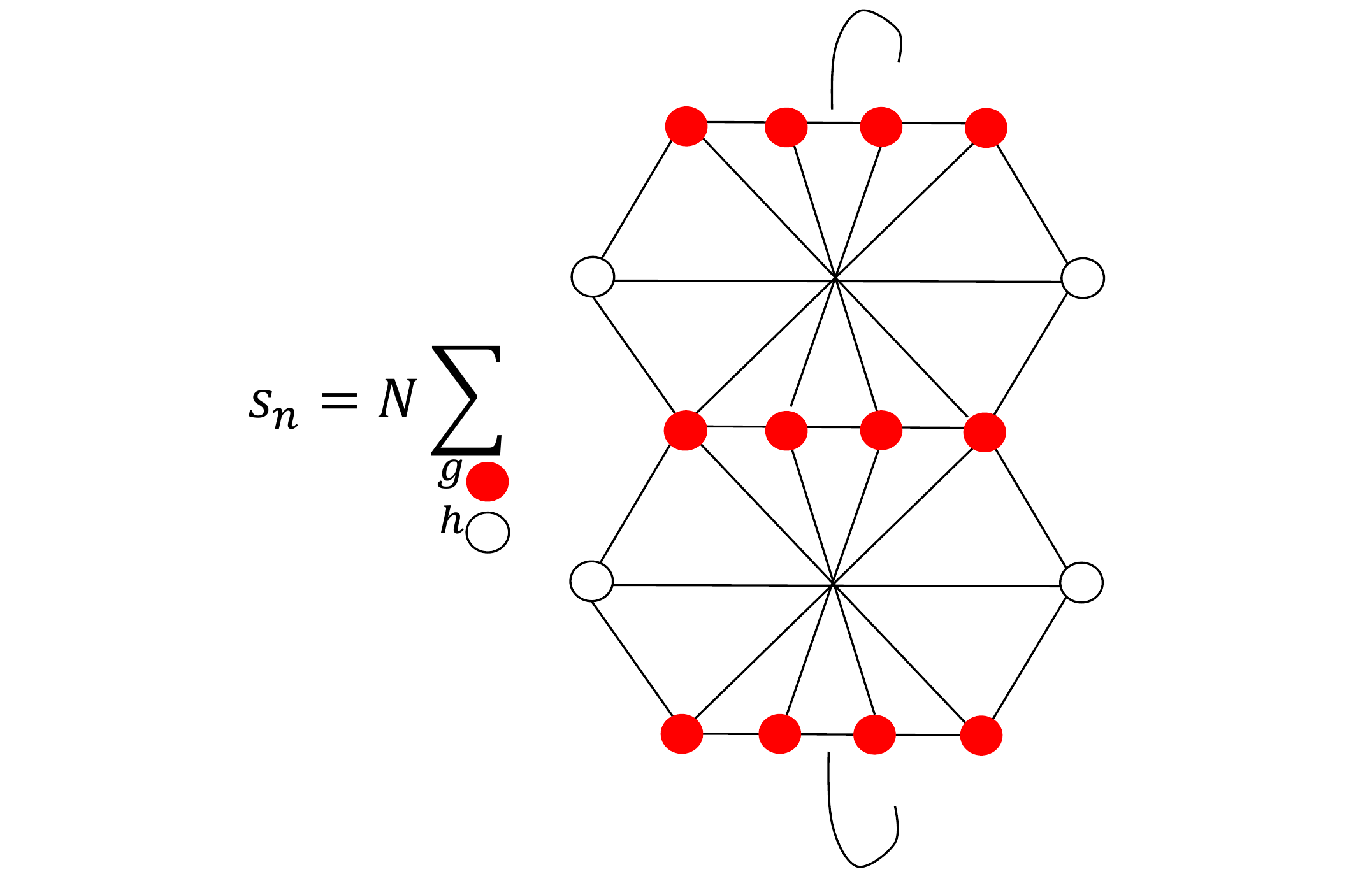}} \nonumber \\
&&\raisebox{-.5\height}{\includegraphics[scale=0.4]{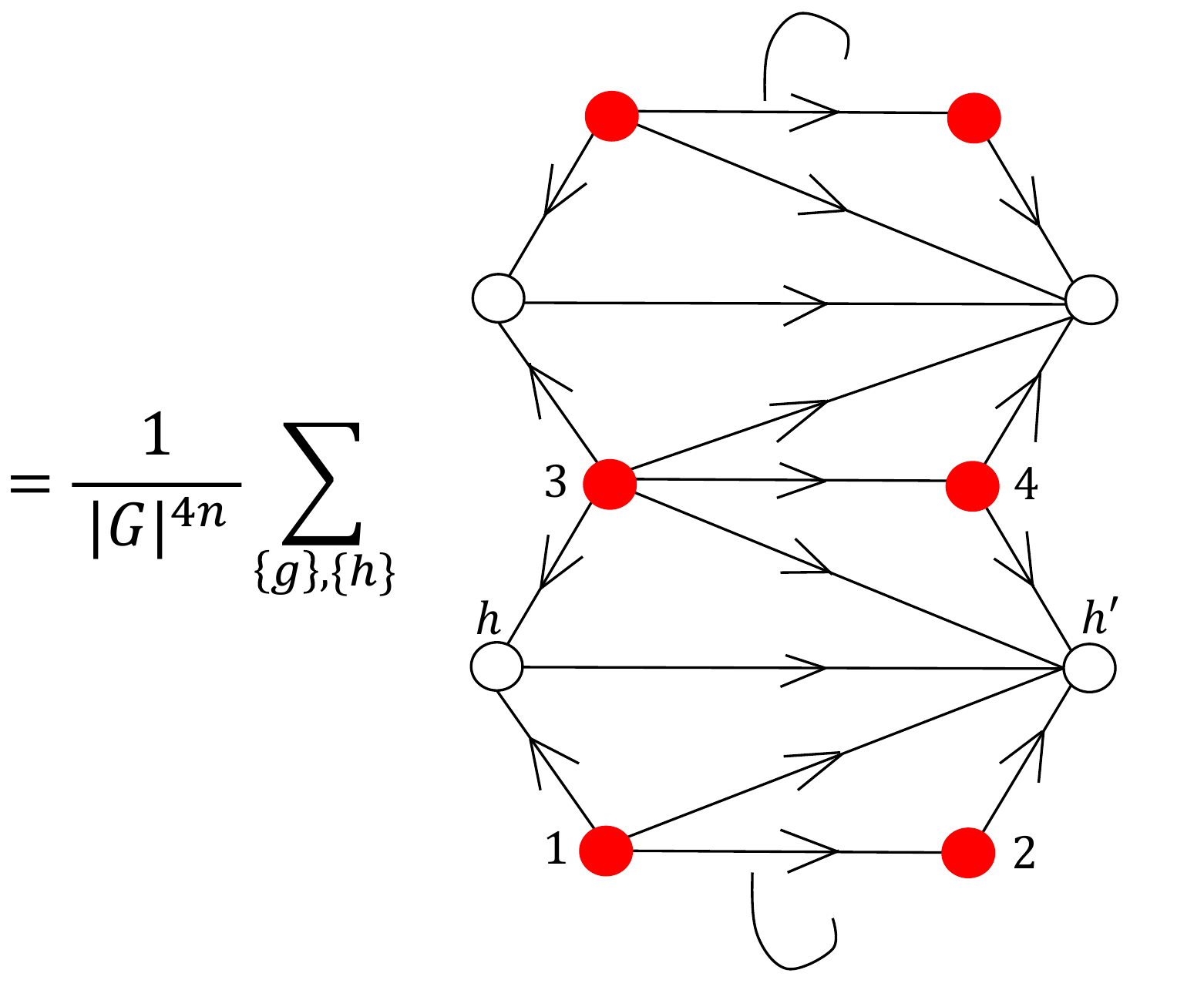}}.
\eea
In the passage to the last complex we used the freedom to remove internal sites, allowing us to leave only two sites in region $A$ in each ``copy", and we also chose a specific branching structure~\cite{chen2011complete}.
Using triangulation invariance we see that it is sufficient to retain two sites in region $A$ (red dots) and two sites in region $B$ (white dots).  %
The resulting $|G|^2 \times |G|^2$ effective density matrix is
\bea
\label{eq:rhoeff}
\rho_{A}^{{\rm{eff}}}(g_1,g_2;g_3,g_4)&& \\=\frac{1}{|G|^4}\sum_{h,h'}&& \frac{\nu_2(g_1,g_2,h')}{\nu_2(g_1,h,h')} \frac{\nu_2(g_3,h,h')}{\nu_2(g_3,g_4,h')}. \nonumber
\eea
It trivially satisfies ${\rm{tr}}\rho_{A}^{{\rm{eff}}}=\sum_{g_1,g_2}\rho_{A}^{{\rm{eff}}}(g_1,g_2;g_1,g_2)=1$.
The SRE can be obtained from
\bea
\label{eq:SRrhoA}
&(U_A(g)\rho_{A}^{{\rm{eff}}})(g_1,g_2;g_3,g_4)=\raisebox{-.5\height}{\includegraphics[width=0.4\linewidth]{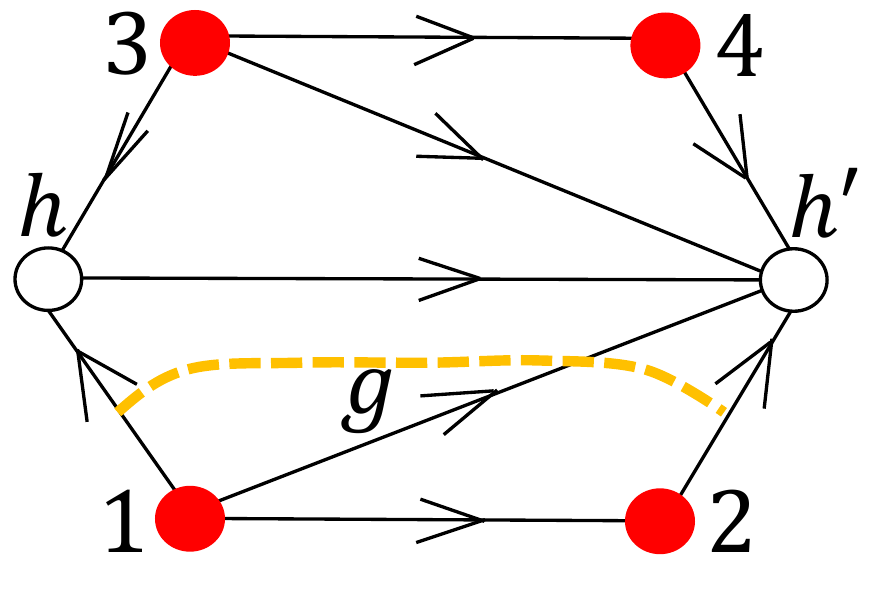}}& \nonumber \\ 
&=\frac{1}{|G|^4}\sum_{h,h'} \frac{\nu_2(gg_1,gg_2,h')}{\nu_2(gg_1,h,h')} \frac{\nu_2(g_3,h,h')}{\nu_2(g_3,g_4,h')}.&
\eea
\subsection{Evaluation of SRE}
Consider the symmetry group $G=\intz_N \times \intz_N$ stabilizing nontrivial SPTs in 1D classified by $\mathcal{H}^{1+d}[G,U(1)]=\intz_N$.~\cite{chen2013symmetry} The case $N=2$ is topologically equivalent to the famous Haldane (or AKLT) topological phase, including also the 1D cluster state. Labeling group elements by a pair of mod-$N$ integers, $g=(n_1,n_2)$ ($n_1,n_2=0, 1,\dots, N-1$), as well as charge sectors $\mathcal{Q}=(q_1,q_2)$  ($q_1,q_2=0, 1,\dots, N-1$), the characters are $\chi_\mathcal{Q}(g)=e^{\frac{2 \pi i}{N} (n_1 q_1+n_2 q_2)}$. Nontrivial cocycles representing the $m$-th phase ($m=0,1,\dots, N-1$) are~\cite{morimoto2014z} 
\be
\label{eq:cocycleznzn}
\nu_2(g_1,g_2,g_3)=e^{\frac{2 \pi i m}{N} [(n_2^2-n_1^2)(n_3^1-n_2^1)]},
\ee where $g_i = (n_i^1,n_i^2)$. 

We can see that $m=0$ is always the trivial phase, with a product state wave function $|\Psi \rangle_{m=0} =\otimes_i \frac{1}{\sqrt{|G|}}\sum_g |g \rangle $. The entanglement spectrum consists then of a single unit eigenvalue in the trivial $\mathcal{Q}=(0,0)$ charge-sector. $m \ne 0$ correspond to topologically nontrivial SPTs. For $\intz_2 \times \intz_2$ the wave function is equivalent to that of the cluster state~\cite{azses2020identification}, and the eigenvalues of the effective density matrix Eq.~(\ref{eq:rhoeff}) are  $\{\lambda_i\}_{\rm{ideal}}=\{1/4, 1/4, 1/4, 1/4, 0, 0, 0, 0, 0, 0, 0, 0, 0, 0, 0, 0\}$. Symmetry resolving those using Eqs.~(\ref{eq:SRrhoA}) and (\ref{charac}), we find that indeed the  eigenvalues are equidecomposed between the 4 symmetry sectors. We note that these ground states exactly correspond to cluster states including higher symmetry generalizations~\cite{zhou2003quantum,miller2016hierarchy,chen2018universal}, see Appendix~\ref{se:groupcoho} for the case of the cluster state.

\begin{figure*}[t]
	\begin{tabular}{ll}
		\includegraphics[width=\linewidth]{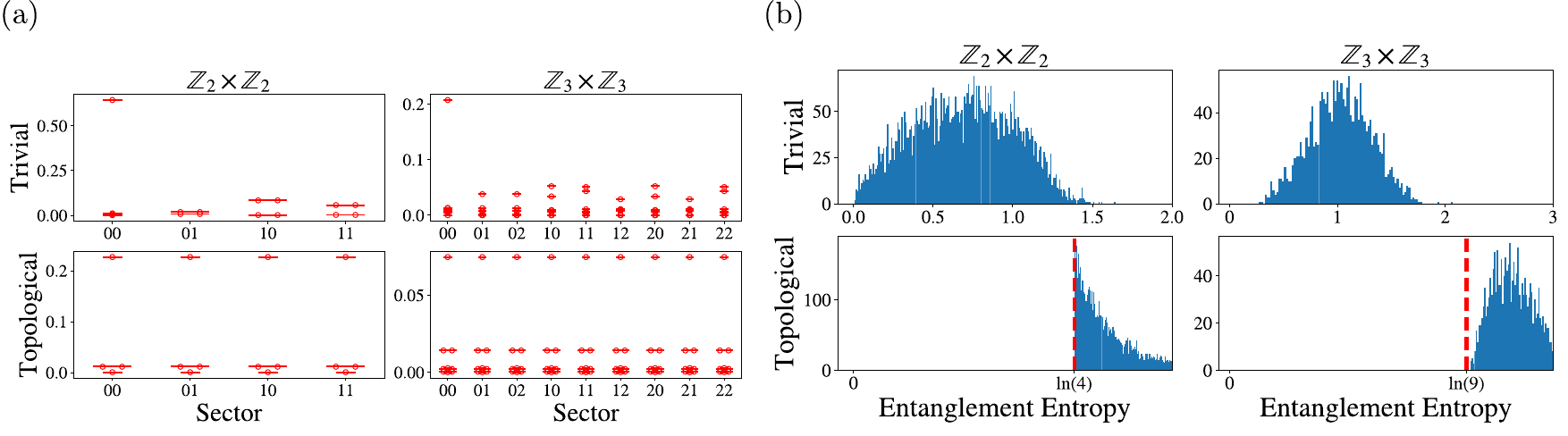}
	\end{tabular}
	\centering
	\caption{(a) Entanglement spectrum for fixed point wave functions given in terms of cocycles Eq.~(\ref{eq:cocycleznzn}) for symmetry group $G=\intz_N \times \intz_N$ with random coboundaries. The topological sectors ($m\ne 0$ in Eq.~(\ref{eq:cocycleznzn})) display degeneracies between the sectors with conserved charge $\mathcal{Q}=(q_1,q_2)$ marked in the $x$-axis. (b) entanglement entropy distribution over a family of wave functions related by coboundary transformations confirming the entanglement minimum in topological phases Eq.~(\ref{eq:Smin}).}
	\label{fig3}
\end{figure*}

The cohomology theory of SPTs allows to explore phases obtained by equivalence classes of the cocycles. We consider cocycles given by Eq.~(\ref{eq:cocycleznzn}) with an extra random coboundary, 
\bea
\nu_2(g_1,g_2,g_3)\to \nu_2(g_1,g_2,g_3) \frac{\nu_1(g_1,g_2) \nu_1(g_2,g_3)}{\nu_1(g_1,g_3)},
\eea 
where $\nu_1(g_1,g_2)=e^{i \theta (g_2 g_1^{-1})}$ and $\theta(g)$ is an arbitrary $g$-dependent angle. Those consist of $|G|$ random variables used to explore each SPT phase. 
In Fig.~\ref{fig3}(a) we plot the SRE spectrum, obtained by diagonalizing the symmetry-resolved density matrix obtained from Eqs.~(\ref{eq:SRrhoA}) and (\ref{charac}), for a specific random coboundary, see Appendix~\ref{se:numerical} for details..  We can see an equidecomposition of the ES into $|G|$ symmetry sectors. This generalizes the case of the pure cocycle Eq.~(\ref{eq:cocycleznzn}) with spectrum $ \{\lambda_i\}_{\rm{ideal}}$
 showing that the content of the ES in each sector is nonuniversal. However the degeneracy always persists in the topological sector. This implies a minimal value of the entanglement entropy: Since there are at least $|G|$ eigenvalues $\le 1/|G|$, the entropy of nontrivial SPTs has a lower bound at
	\be
	\label{eq:Smin}
S_{\mathrm{non~trivial~SPT}} \ge \log |G|,~~~{\rm{(equidecomposition)}}
\ee 
as illustrated by the dashed lines in the histogram plots in Fig.~\ref{fig3}(b). On the other hand  trivial SPTs can be arbitrarily close to product states (although statistically they are typically not) and have no topologically protected entanglement minimum.

\section{Proof of equidecomposition}
As our numerical results exemplify, the SRE spectrum and the entanglement entropy are not universal quantities of SPT phases, i.e., these quantities vary within phases. %
However the equidecomposition is a universal property of nontrivial SPTs. In this section we prove this analytically for finite Abelian groups.

Consider the quantity
\be
Z_n(g) \equiv {\rm{Tr}} U_A(g) (\rho_A^{{\rm{eff}}})^n.
\ee
It has the graphical representation  of a partition function on an $n$-sheet Riemann surface as in Fig.~\ref{fig1}(d), with an additional defect line. We will show that 
\be
\label{condition}
Z_n(g)=0~{\rm{for}} ~ g \ne e
\ee
holds in nontrivial SPTs, independent of coboundary transformations. Here $e \in G$ is the identity element. Namely,  the topological path integral  vanishes in the topological phase when inserting a nontrivial defect line. Combined with Eq.~(\ref{charac}), we have
\be
s_n(\mathcal{Q})=\frac{1}{|G|} \sum_{g \in G} \chi_\mathcal{Q}(g) Z_n(g)=\frac{1}{|G|} \chi_\mathcal{Q}(e) Z_n(e)=\frac{1}{|G|} s_n,
\ee
which is independent of $\mathcal{Q}$, so that equidecomposition follows. In the rest of this section we turn to a proof of Eq.~(\ref{condition}).

First we will prove it algebraically for $n=1$. Then, we will provide a geometric interpretation to this proof in terms of topological path integrals and using their triangulation invariance properties, allowing to generalize the proof for any $n$.

\subsection{$n=1$: Symmetry-resolved probabilities}
\label{se:n1}
\begin{figure*}[t]
	\centering
	\includegraphics[scale=0.5]{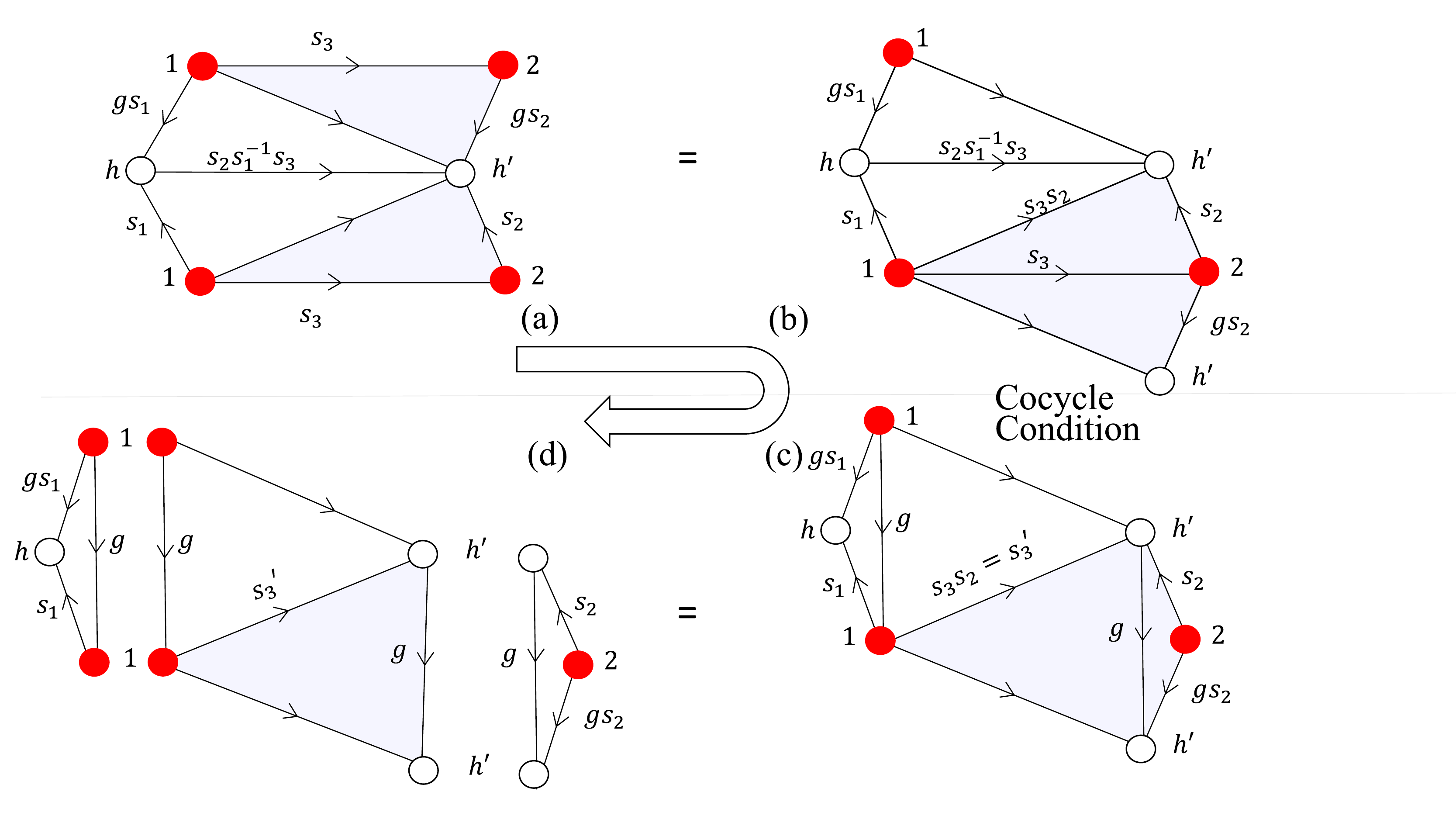}
	\caption{Graphical representation of the calculation of the symmetry-resolved first \Renyi entropy $(U_A(g)\rho_A^{\rm{eff}})(g_1,g_2;g_3,g_4)$ in Sec.~\ref{se:n1}.}
	\label{fig4}
\end{figure*}
\begin{figure*}[t]
	\centering
	\includegraphics[scale=0.5]{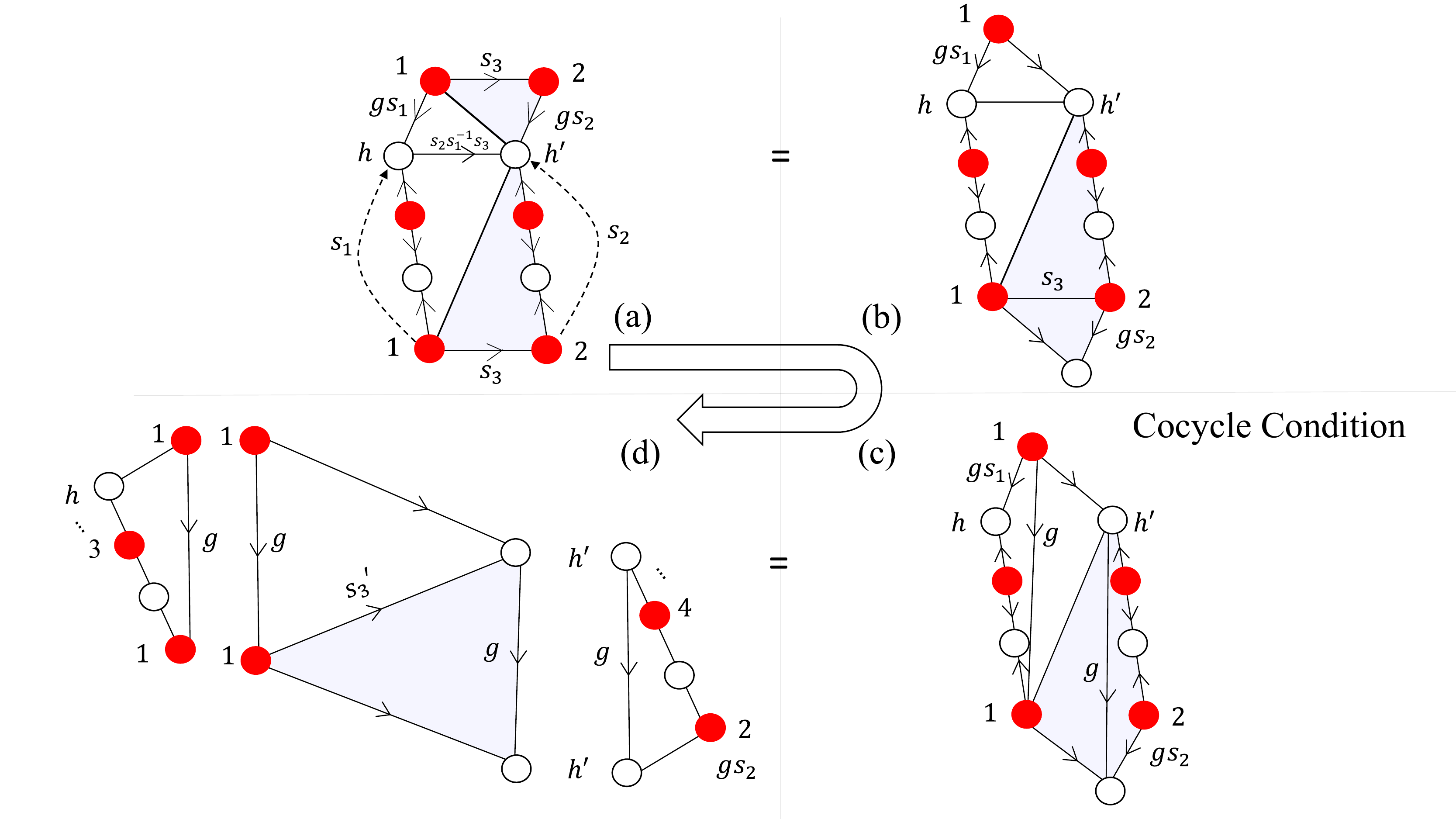}
	\caption{Generalization of Fig.~4 for arbitrary $n$.}
	\label{fig5}
\end{figure*}

We begin with the effective density matrix of the sector $g$, $(U_A(g)\rho_A^{\rm{eff}})(g_1,g_2;g_3,g_4)$ given in Eq.~(\ref{eq:SRrhoA}). We consider $Z_g  \equiv \Tr(U_A(g)\rho_A^{\rm{eff}})$.
The trace %
is given by (after applying a coboundary $\nu_1$)
\begin{widetext}
\be
\label{eq:Zg}
Z_g=\frac{1}{|G|^4}\sum_{g_1,g_2,h,h'} \frac{\nu_2(gg_1,gg_2,h')}{\nu_2(gg_1,h,h')} \frac{\nu_2(g_1,h,h')}{\nu_2(g_1,g_2,h')} \frac{\nu_1(gg_2,h')\nu_1(g_1,h)}{\nu_1(g_2,h')\nu_1(gg_1,h)}.
\ee
We write the 3-variable $\nu_2$'s in terms of 2-variable $\omega$'s, $\nu_2(g_1,g_2,g_3)=\omega_2(g_1^{-1}g_2,g_2^{-1}g_3)$, so that $\nu$ automatically satisfies Eq.~(\ref{eq:symcocycle})~\cite{chen2013symmetry}. This yields (we denote $\omega_1(g) = \beta(g)$)
$$Z_g =\frac{1}{|G|^4}\sum_{g_1,g_2,h,h'} \frac{\omega_2((gg_1)^{-1}gg_2,(gg_2)^{-1}h')}{\omega_2((gg_1)^{-1}h,h^{-1}h')} \frac{\omega_2(g_1^{-1}h,h^{-1}h')}{\omega_2(g_1^{-1}g_2,g_2^{-1}h')} \frac{\beta((gg_2)^{-1}h')\beta(g_1^{-1}h)}{\beta(g_2^{-1}h')\beta((gg_1)^{-1}h)}.$$
Recalling that the group is Abelian, let us define $s_1=g^{-1} g_1^{-1}h$, $s_2=g^{-1} g_2^{-1} h'$, and $s_3=g_1^{-1}g_2$. These variables live on the links of the complex, see Fig.~\ref{fig4}(a). Since the $\nu$ cocycles satisfy the symmetry condition, after the change of variables the ``center of mass" sum simply yields a factor $|G|$, $\sum_{g_1,g_2,h,h'}=|G| \sum_{s_1,s_2,s_3}$, and we obtain
$$Z_g=\frac{1}{|G|^3}\sum_{s_1,s_2,s_3} \frac{\omega_2(s_3,s_2)}{\omega_2(s_3,gs_2)} \frac{\omega_2(gs_1, s_2s_1^{-1}s_3)}{\omega_2(s_1,s_2s_1^{-1}s_3)}  \frac{\beta(s_2) \beta(gs_1)}{\beta(s_1)\beta(gs_2)}.$$
One can see in Fig.~\ref{fig4}(a) that indeed the arguments of the coboundaries $\beta$ appear on the boundary of the complex.
Separating $s_1,s_2,s_3$ is done by using the cocycle conditions
\be
\label{cpcy}
\frac{\omega_2(s_3,s_2)}{\omega_2(s_3,s_2g)}=\frac{\omega_2(s_2,g)}{\omega_2(s_3s_2,g)},~~~\frac{\omega_2(gs_1,(s_2s_1^{-1}s_3))}{\omega_2(s_1,(s_2s_1^{-1}s_3))}=\frac{\omega_2(g,s_1(s_2s_1^{-1}s_3))}{\omega_2(g,s_1)}=\frac{\omega_2(g,s_2s_3)}{\omega_2(g,s_1)},
\ee
and then changing the sum variables defining $s_3' = s_3 s_2 \to s_3$ obtaining
\be
\label{resiltinter}
Z_g=\frac{1}{|G|^3} \left[ \sum_{s_1} \frac{\beta(s_1g)}{\omega_2(g,s_1)\beta_1(s_1)} \right] \left[ \sum_{s_2} \frac{\omega_2(s_2,g)\beta(s_2)}{\beta(s_2g)} \right] \left[ \sum_{s_3} \frac{\omega_2(g,s_3)}{\omega_2(s_3,g)} \right].
\ee
\end{widetext}
This is the required form. We managed to separate the sum of products, into 3 products of sums. Only the  sum over $s_3$ is coboundary-independent and so in general only when this sum vanishes for $g\neq e$ we will have equidecomposition. We demonstrate that 
\be
\label{DanielEq}
 \sum_{s_3} \frac{\omega_2(g,s_3)}{\omega_2(s_3,g)} =0, ~~~~(g \ne e)
\ee 
in the Appedix~\ref{se:prrofcondition} using group theory methods. We note that this does not follow from cocycle conditions, but rather, by properties of Abelian cocycles. Essentially, this sum has the strucure of a geometric series of phases and hence is finite only in the trivial cocycle. 

\subsection{Graphical generalization}
\label{se:generaln}
The steps involved in the algebraic calculation of the preceding subsection can be graphically represented as in  Fig.~\ref{fig4}(a-d). The crucial step is the use of the cocycle condition Eq.~(\ref{cpcy}), represented by the transition Fig.~\ref{fig4}(b) $\to$ (c). We see that through this algebraically allowed step, we have connected a vertex to itself, via the $g-$link. In Fig.~\ref{fig4}(d) we observe that the summations over products of cocycles factorizes, as in Eq.~(\ref{resiltinter}). 

Now consider $ {\rm{Tr}} U_A(g) (\rho_A^{{\rm{eff}}})^n$. As in Eq.~(\ref{eq:sn2}) it is written in terms of the sum $\frac{1}{|G|^{4n}} \sum_{g,h}$ of a large complex. According to the main Dijkgraaf-Witten theorem (theorem 2 in Ref.~\cite{mesaros2013classification}) this depends only on the triangulation and the values of $\{g_i,h_i \}$ on the boundary. In Fig.~5(a), we start from a different triangulation than in Eq.~(\ref{eq:sn2}), that connects sites far-away in replica space ($n$), This  is convenient since it allows us to repeat the same calculation we did for $n=1$. Following the same steps, shown in Fig.~\ref{fig5}(a-d) for general $n$ we obtain a similar factorization of the complex, where one of the factors is coboundary-independent as well as $n$-independent, and vanishes for the topological-nontrivial cocycles,
\be
{\rm{Tr}} U_A(g) (\rho_A^{{\rm{eff}}})^n \propto     \sum_{s_3} \frac{\omega_2(g,s_3)}{\omega_2(s_3,g)} =0, ~~~~g \ne e.
\ee
This, together with the demonstration of Eq.~(\ref{DanielEq}) given in Appendix~\ref{se:prrofcondition}, completes our proof. In the next section we work out more examples.

\section{Entanglement equidecomposition in finite Abelian groups}
We  defined the resolution of the entanglement spectrum into symmetry sectors $\{\lambda_i \}_\mathcal{Q}$ and found equidecomposition for certain examples. Here we provide a general condition for equidecomposition in finite Abelian groups using a general form of the cocycles~\cite{berkovich1998characters}. %
For a finite Abelian group $G$, there is always a decomposition such that $G = \intz_{e_{1}} \times \intz_{e_{2}} \times \dots \times \intz_{e_{k}}$, where $e_{i}$ divides $e_{i+1}$. Group elements are $\{g_1,\dots, g_k \}$ where $g_i \in \intz_{e_i}$. It is possible to enumerate all the cocycles, and hence all the SPTs, using a set of integers $p_{ij}$, ($1 \le i,j \le k$) where $0 \le p_{i <j} < gcd(e_i,e_k)$ and $p_{i \geq j}=0$.  We find that if for all group elements $g \neq e$, there exists $r$ ($1 \le r \le k$) such that
\be
\label{condequi}
\sum_{i=1}^k  \frac{(p_{ri}-p_{ir})g_i}{{\rm{min}}(e_i,e_r)} \notin \intz,
\ee
then there is equidecomposition. %
In Appendix~\ref{se:prrofcondition} we show that this condition guarantees Eq.~(\ref{DanielEq}). As we proved in Sec.~\ref{se:generaln},   Eq.~(\ref{DanielEq}) guarantees equidecomposition of the symmetry-resolved $n-$\Renyi entropy for any $n$. This implies a degeneracy in the entanglement spectrum. Furthermore, in Appendix~\ref{se:eqnmc} we show that this condition is equivalent to the concept of maximally non-commutative (MNC) cocycles, establishing a connection between equidecomposition in the entanglement spectrum and these MNC phases that are known to allow measurement-based quantum computation ~\cite{else2012symmetry,de2020inaccessible,stephen2017computational}.

As will be discussed in the examples below, for some groups  condition~(\ref{condequi}) holds for almost but not for all $g \neq e$, and then we find that most, but not all of the symmetry sectors, are degenerate.

\subsubsection{$\intz_N \times \intz_N$}
Let us focus on the group $\intz_N \times \intz_N$. The aforementioned decomposition of this group is given by $e_1=e_2=N$ with $k=2$. Denoting $m=p_{12}$,  condition~(\ref{condequi}) then reads: For all $g\neq e$, $  \frac{mg_1}{N} \notin \intz$ or $\frac{mg_2}{N} \notin \intz.$

For the trivial phase $m=0$, the condition never holds as $0 \in \intz$, and so equidecomposition does not occur. On the contrary, for non-trivial phase $p_{12} \neq 0$, for $g\neq e$ either $g_1$ or $g_2$ is nonzero. Therefore, for this component $g_i$ we have that $N$ does not divide $mg_i$ for all $g_i$ in case of equidecomposition. This is possible if and only if $\mathrm{gcd}(m,N)=1$. Specifically, for prime $N$ equidecomposition always occurs.

Using numerical simulations we now check our condition and also test further implications. Specifically we test cases with special symmetry groups where Eq.~(\ref{condequi}) holds for almost but not all group elements, leading to a degeneracy between a subset of symmetry sectors. We compute $\{ Z_g \}$ for all $g \in G$, as defined in Eq.~(\ref{eq:Zg}) using the $\intz_N \times \intz_N$ cocycles, with random coboundaries, and then use the $\intz_N \times \intz_N$ characters in order to obtain the symmetry-resolved probabilities $\{ Z_\mathcal{Q} \}$ using Eq.~(\ref{charac}); for further details see Appendix~\ref{se:numerical}. In Table~\ref{tb:num_zn} we plot the different sectors' ``partitions", i.e. the number of different values among the $N^2$ probabilities $\{ Z_\mathcal{Q} \}$. We made sure that nonuniversal degeneracies are removed using random coboundaries. We indeed see that for prime $N$ and for non-trivial phases there is always full equidecomposition, i.e., one partition. Moreover, when
$m$, indexing the topological sector, divides $N$, we see that although equidecomposition does not occur, we have various sectors with degenerate eigenvalues, and so we see ``almost" equidecomposition. These patterns, as well the complicated ones, are well understood by employing condition~(\ref{condequi}).

While both full equidecomposition for $N$ prime or partial degeneracy occurring when $m$ divides $N$, are signatures of topological nontrivial phases, Table~\ref{tb:num_zn} also displays degeneracies in the trivial phases for $N>2$ in the form of $(2,2,....,2,1,1,\dots,1)$. Generally, and including in the trivial phase, there is a degeneracy in the entanglement spectrum between charge sectors $\mathcal{Q}$ and $\mathcal{Q}^{-1}$. This fact can be also seen in Fig.~\ref{fig3}(a) for $G=\intz_3 \times \intz_3$ specifically in the trivial phase. This follows from the relation $Z_g = (Z_{g^{-1}})^*$, which can be proven by gluing together two discs of Fig.~\ref{fig2}(a) with opposite orientations, obtaining a closed surface ($S^2$) with a closed defect. From here, it is clear that the symmetry sectors will have $(2,2,....,2,1,1,\dots,1)$ pattern as the characters for $\intz_N \times \intz_N$ obey similar relations $(\chi_q(g)= \chi_q(g^{-1})^*)$. The  double degeneracies  (2's) come from group elements $g$ such that $g \neq g^{-1}$, and the (1's) come from elements  $g=g^{-1}$.

\begin{table}
	\begin{ruledtabular}
		\begin{tabular}{cccc}
			$N$ & $m$ & Partitions & Signature \\
			2 & 0 & 4 & (1,1,1,1) \\
			2 & 1 & 1 & (4) \\
			\hline
			3 & 0 & 5 & (2,2,2,2,1) \\
			3 & 1 & 1 & (9) \\
			3 & 2 & 1 & (9) \\
			\hline
			4 & 0 & 10 & (2,2,2,2,2,2,1,1,1,1) \\
			4 & 1 & 1 & (16) \\
			4 & 2 & 4 & (4,4,4,4) \\
			4 & 3 & 1 & (16) \\
			\hline
			5 & 0 & 13 & (2,2,\dots,2,2,1) \\
			5 & 1 & 1 & (25) \\
			5 & 2 & 1 & (25) \\
			5 & 3 & 1 & (25) \\
			5 & 4 & 1 & (25) \\
			\hline
			6 & 0 & 20 & (2,2,\dots,2,2,1,1,1,1) \\
			6 & 1 & 1 & (36) \\
			6 & 2 & 4 & (9,9,9,9) \\
			6 & 3 & 5 & (8,8,8,8,4) \\
			6 & 4 & 4 & (9,9,9,9) \\
			6 & 5 & 1 & (36) \\
		\end{tabular}
	\end{ruledtabular}
	\caption{\label{tb:num_zn} Numerical results for the sectors' partition (see text for definition) of the group $\intz_N \times \intz_N$. The index $m$ labels topological phases. Partitions are the number of different sectors. Signature shows the different sectors (there are $N^2$ sectors) partitioned by their equality (e.g. $(1,1,1,1)$ means that all the sectors are different while $(4)$ means they are all equal and $(2,2)$ means that there are $2$ pairs of $2$ equal sectors).}
\end{table}

\begin{figure*}[t]
	\centering
	\includegraphics[scale=0.4]{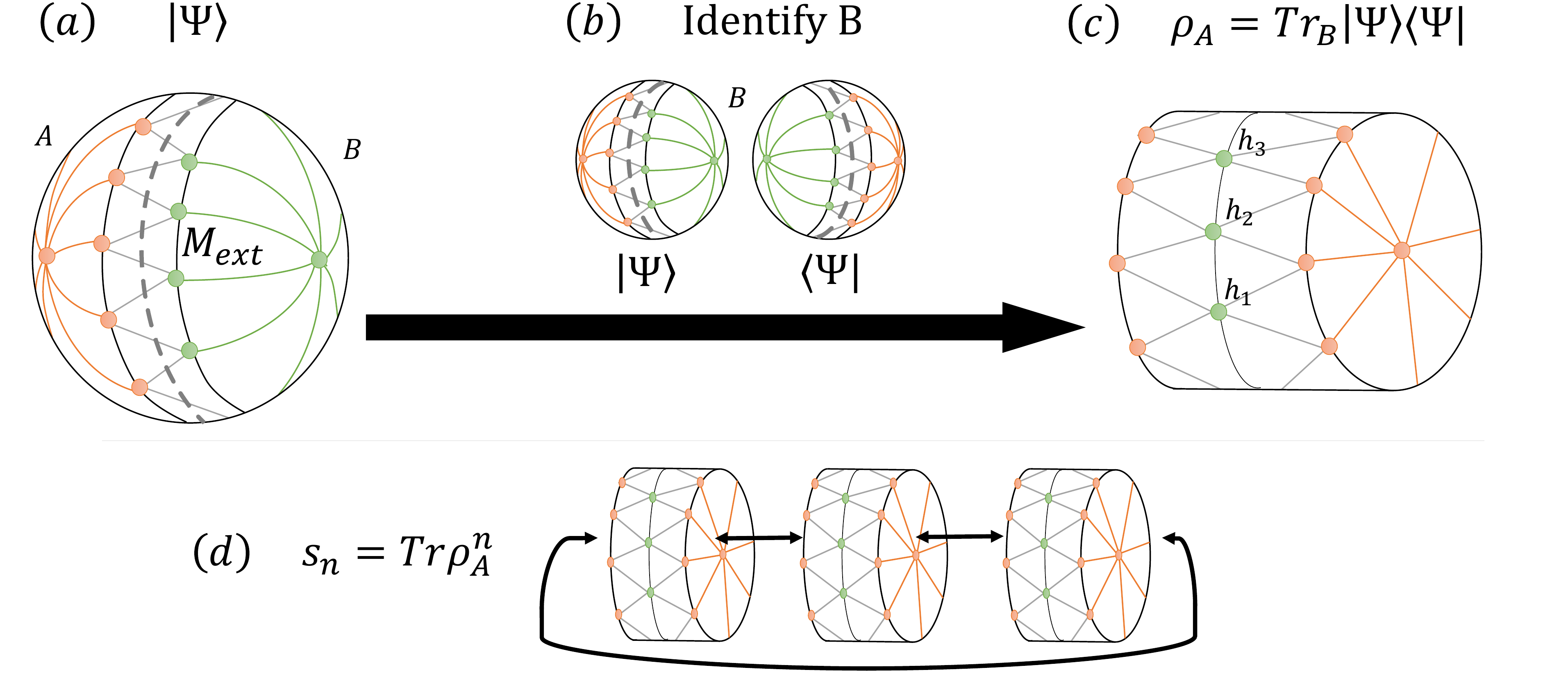}
	\caption{(a) A 2D SPT wave function on a triangulated 2-sphere, which is partitioned into regions $A$ and $B$. The wave function is obtained from the action amplitude in Eq.~4 on the extended manifold $M_{ext}$ corresponding to the 3-ball.  Gluing the $B$ regions together of two such spheres of opposite orientation (b), yields the reduced density matrix $\rho_A$ (c).   $\rho_A$ contains $L$ boundary points $(h_i)$ $(i=1,\dots,L)$ which can not be removed by triangulation. (d) The $n$-th \Renyi entropy of a 2D SPT state corresponds to a solid torus having $L \cdot 2n$ points $((h_i)_k)$ $(i=1,\dots,L,~k=1,\dots,n)$ from $B$ and corresponding points from $A$ on its boundary (not shown).}
	\label{fig6}
\end{figure*}

\subsubsection{$\intz_2^n$}

In the case of $G = \intz_2^n$, we have $\intz_2^{\frac{n(n-1)}{2}}$ phases (from the cohomology group)~\cite{mesaros2013classification}. The phases can be represented by strictly upper triangular $n \times n$ matrix $A$ with entries in $\intz_2$. For the usual cocycles with no coboundary, we can calculate the sectors from the common kernel of $A,A^T$. This common kernel is hard to calculate generally but unveils the degeneracy patterns when using general principles in group theory.

Let us use the general formula Eq.~(\ref{condequi}) to calculate the cases in which equidecomposition occurs. The condition reads: For all $g \neq e$ there exists $r$ such that $\sum_i (A_{ri}-A_{ir})g_i = 1$, or equivalently, $A\ket{g}\neq A^T\ket{g}$ when using addition and multiplication in $\intz_2$, where $\ket{g}$ is a vector with $n$ components in $\intz_2$ representing an element in $G$.

For the trivial phase $A=0$, it is clear that equidecomposition never occurs as $A=A^T$. For the non-trivial phases in the case that $k$ is even, specifically, where $A$ is a matrix with $1$'s on the anti-diagonal ($a_{i,j}$ with $i+j=n+1$) above the main diagonal, we have that indeed this condition for equidecomposition holds as one can check ($A-A^T$ is invertible, as the determinant is trivially non-zero, and hence the only vector obeying $A-A^T \ket{v}=0$ is the zero vector, which is $e$). For $n$ odd that is not the case, as $B=A-A^T$ is a skew-symmetric matrix obeying $B=-B^T$, which implies that the determinant vanishes for odd sized matrices, and there is no equidecomposition for any phase.
 
\section{Prospect for higher dimensional generalizations} 
SPT phases in spatial dimensions $d>1$ are much less understood compared to their one-dimensional counterparts. In 2D it is believed that the cohomological picture still provides a complete classification. The fate of the edge states is more involved: For a model with on-site discrete symmetry Chen \emph{et. al.}~\cite{chen2011two} and Levin and Gu~\cite{levin2012braiding} showed that the 1D system on the boundary must be gapless if the symmetry is not  broken. Also long-range entangled gapless states may show nontrivial boundary states protected by symmetry~\cite{scaffidi2017gapless}. A useful direction to higher dimensions relies on observing the structure of domain walls~\cite{chen2014symmetry} which  can carry topological order of lower dimension. While cohomology theory is complete in 1D, in which it is equivalent to the classification of projective representations of the symmetry, and agrees with other classification approaches in 2D~\cite{lu2012theory}, beyond-cohomology approaches have been proposed in higher dimensions implying that the cohomology classification is incomplete in general. The entanglement spectrum of the Levin and Gu model in 2D was studied~\cite{zaletel2014detecting} revealing unique signatures of the topological phase. The gapless edge states results in a more involved and gapless entanglement spectrum, which nevertheless admits a symmetry decomposition. A more general study of entanglement, and specifically SRE of higher dimensional SPTs can shed light on these phases of matter.

In this section we outline how our geometric approach can be used to compute the SRE of higher dimensional wave functions constructed using cohomology. Consider a generalization of Fig.~1 to a 2D phase. In this case, the wave function of a $d=2$ SPT on a 2-sphere $M=S^2$ is obtained as the action amplitude Eq.~(\ref{eq:WENwf}) on the extended manifold $M_{ext}$ being a 3-ball, $M=\partial M_{ext}$. A minimal triangulation of this 3D object using tetrahedras involves a 2D triangulation of the surface $S^2$ and a single point inside $M_{ext}$. In Fig.~\ref{fig6}(a) we display a triangulated SPT on $S^2$ which is bipartitioned into two half-spheres yielding a 1D boundary carrying the entanglement between $A$ and $B$. We now construct the reduced density matrix $\rho_A=\Tr_B |\Psi \rangle \langle \Psi |$. We glue together two such spheres with opposite orientations (corresponding to $|\Psi \rangle$ and $\langle \Psi |$). As can be seen in Fig.~\ref{fig6}(c), the resulting object has a pair of boundaries on the left and right corresponding to $|\Psi_A \rangle$ and $ \langle \Psi_A |$, as well as a 1D boundary denoted $h_i$ which is a reminiscent of $B$. While in the 1D case discussed throughout the paper we had only a pair of points $h,h'$, now we have $L$ such points where $L$ is the length  of the entanglement cut. Now the $n-$th \Renyi entropy is obtained by periodically gluing $n$ such objects,  Fig.~\ref{fig6}(d). The resulting manifold has $L \cdot 2n$ points ($(h_i)_{k=1,\dots , n}$ from $B$, and similarly for $A$) that can not be removed by retriangulation. As a result, the computation of $s_n=\Tr[\rho_A^n]$ for the cohomological wave function involves a sum of order of $L \cdot n$ terms, each of which consists of a product of an order of $L \cdot n$ 3-cocycles. The SRE can be computed by an insertion of a 2-membrane defect into the interior of the object in Fig.~\ref{fig6}(d). The boundary of this defect coincides with the boundary of $A$, $\partial D= \partial A$, which on the triangulated space consists of a circle separating the points originating from one side of $A$ from those of $B$ at an arbitrary $k$.

Probing the gapless entanglement spectrum requires to capture the thermodynamic limit $L \to \infty$.  This can be achieved using a transfer martix approach, where the \Renyi entropy takes the form $s_n \sim \rm{Tr}[A^L]$~\cite{TranferInPrep}. The size of the transfer matrix is $\sim |G|^{2n}$. As an example, one can compute the first few \Renyi entropies for the Levin-Gu model~\cite{TranferInPrep}.
 
\section{Conclusions}
We employed the Dijkgraaf-Witten~\cite{dijkgraaf1990topological} discrete gauge theories and the associated cohomological classification of symmetry-protected topological order, in order to describe the entanglement structure of SPTs. These gauge theories are based on terms which are topological invariants of closed manifolds. %
We showed  in general dimensions that entanglement measures are represented by generalized $n$-sheet Riemann surfaces which have a boundary; as a result entanglement itself is not a topological invariant. However, one expects to find topologically invariant features in the entanglement of SPTs. For this purpose we employed symmetry-resolved entanglement. Focusing on 1D, we found generally that in topological phases, the reduced density matrix decomposes into identical blocks. The probabilities to be in various symmetry sectors are then identical. Our equidecomposition of the reduced density matrix into identical blocks is directly  connected to the known degeneracies in the entanglement spectrum.

We provided a proof of the absolute entanglement equidecomposition using the underlying cohomological description for general SPTs stabilized by Abelian fintie symmetry groups. This leads to a minimum entanglement of a given phase. Also, some symmetry groups contain phases in which equidecomposition is replaced by a partial degeneracy between symmetry sectors, like $G=\intz_N \times \intz_N$ with $N$ non-prime. Yet, we did not find a formulation of these degeneracies in terms of a Dijkgraaf-Witten path integral over a closed manifold. Rather, it appears to be a special property of cocycles of Abelian groups which we considered. 

In 1D, similar results can be obtained somewhat more easily from matrix product state (MPS) considerations~\cite{pollmann2010entanglement,de2020inaccessible}. Our approach, however, offers some hope of generalization to higher dimensions. On the other hand, MPS arguments seems at first sight to be more general since any gapped ground state in 1D can be accurately represented by an MPS with some finite bond dimension, so that the entanglement spectrum mirrors a real edge spectrum with edge modes giving rise to degeneracies~\cite{pollmann2010entanglement,de2020inaccessible}. In contrast, our results apply for the wave functions constructed by Chen et al.~\cite{chen2011classification} describing fixed point states with zero correlation length, and thus to parent Hamiltonians of a particular form. Nevertheless, the simplicity of these states together with the mathematical toolbox of cohomology can then open the way to explore using our methods the entanglement structure of other systems including higher dimensional generalizations~\cite{chen2011two,levin2012braiding,zaletel2014detecting,scaffidi2017gapless,chen2014symmetry}, which is left for future work.

\section{Acknowledgements}
We acknowledge useful discussions with Emanuele G. Dalla Torre, Moshe Goldstein, and Ari Turner. We thank support from ARO (W911NF-20-1-0013) and the Israel Science Foundation grant number 154/19. DA acknowledges the Erasmus+ programme of the European Union grant and hospitality of the Institute for Quantum Optics and Quantum Information at the University of Innsbruck during which this work was finalized.

\newpage

\appendix*
\section{}
\appendix

In Appendix~\ref{se:groupcoho} we review the definitions of Ref.~\cite{chen2013symmetry} for group cocycles, their graphical representation on simplexes and complexes, and their use application to write ideal SPT wave functions. We also define partition functions with defects used to compute symmetry-resolved entanglement. In Appendix~\ref{se:prrofcondition} we use known forms of cocycles of general finite Abelian groups to prove Eqs.~(\ref{DanielEq}) and (\ref{condequi}) which guarantee equidecomposition of entanglement. In Appendix~\ref{se:numerical} we provide details on our numerical simulations. In Appendix~\ref{se:eqnmc} we prove a general relation between the degeneracies in the entanglement spectrum and the non-commutativity of the cocycles, specifically, we show that equidecomposition and maximally non-commutative cocycles are equivalent.

\section{Geometrical interpretation of group cohomology}
\label{se:groupcoho}

 \subsection{Cochains, cocycles, coboundaries and the cohomology group}
 A $d$-cochain of a group $G$ is an arbitrary complex function $\mu_d(g_0,g_1, \dots , g_d)$ of $d+1$ $G$-valued variables satisfying $|\mu_d(g_0,g_1, \dots , g_d)|=1$ and 
 \be
 \label{cochain}
 \mu_d(g_0,g_1, \dots , g_d)=\mu_d(g g_0,g g_1, \dots , g g_d),~~~~g \in G.
 \ee
 A $d$-cocycle is a special $d-$cochain that satisfies $\prod_i \nu_d^{(-1)^i}(g_0,\dots g_{i-1}, g_{i+1}, \dots , g_d)=1$. As a specific examples see Eqs.~(\ref{1cocycle}) or (\ref{2cocyclee}) below.
 
 A $d-$coboundary $\lambda_d$ is a special $d-$cocycle constructed from $(d-1)$-cochains $\mu_{d-1}$, $\lambda_d(g_0, \dots, g_d)=\prod_{i=0}^d \mu_{d-1}^{(-1)^i}(g_0, \dots, g_{i-1},g_{i+1}, \dots, g_d)$. As a specific examples see Eqs.~(\ref{2cob}) or (\ref{3cob}) below.
 
 Two cocycles are  equivalent if they differ by a coboundary. Equivalence classes of cocycles are given by the $d-$cohomology group $\mathcal{H}^d[G,U(1)]$ of the group $G$.
 
 \subsection{Example: $d=1$, $G=\intz_N \times \intz_N$}
 The d-cochain $\mu_d(g_0,g_1, \dots , g_f)=1$ for all $g's$ is  a trivial cocycle. As a specific but central example for  nontrivial cocycles, consider the symmetry group $G=\intz_N \times \intz_N$ for which $\mathcal{H}^2[G,U(1)]=\intz_N$. Thus there are $N$ equivalence classes.  The $m-th$ cocycle is
 \be
 \nu_2(g_1,g_2,g_3)=e^{\frac{2 \pi i m}{N} [(n_2^2-n_1^2)(n_3^1-n_2^1)]}
 \ee
 where $n_{i}^j = 0, 1,\dots ,N-1$ is the $j$-th component of $g_i$.
 
 \subsubsection{Relation to cluster states for $G=\intz_2 \times \intz_2$}
 The states with $m=0$ and $m=1$ exactly map to the ground states of the Hamiltonians
 \bea
 \label{eq:exactcocyclezn}
 H_{m=0}&=&-\sum_i X_i,\nonumber \\
  H_{m=1}&=&-\sum_i Z_{i-1} X_i Z_{i+1},
 \eea
 where $X_i, Z_i$ are Pauli matrices acting on site $i$.
 To see the relation to the basis used in the cohomology description one pairs up neighboring sites $(2i,2i+1)$ and works in the $Z$ basis and defines for this effective site $g_i= \{ n_i^1, n_i^2 \} = \{\frac{1-Z_{2i}}{2}, \frac{1-Z_{2i+1}}{2} \}$~\cite{else2012symmetry}. For example, the ground state of $H_{m=0}$ is the product state $\otimes_i (|+\rangle_{2i} |+\rangle_{2i+1})$, which the same as $\frac{1}{\sqrt{|G|}}\sum_g |g\rangle$, consistent with Eq.~(\ref{eq:exactcocyclezn}) for $m=0$. 
 \subsection{Graphical representation}
 The above definitions admit a useful graphical representation. For $d=1$, $\nu_1(g_1,g_2)$ is a 1-cocycle if 
 \be
 \label{1cocycle}
 ~\frac{\nu_1(g_1,g_2)\nu_1(g_0,g_1)}{\nu_1(g_0,g_2)}=1=~\raisebox{-.5\height}{\includegraphics[scale=0.5]{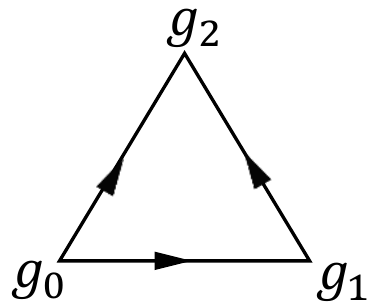}},
 \ee
 for any $g_0,g_1,g_2 \in G$.
 The diagram represents a 1D complex, composed of three  1D simplexes (lines), each of which corresponds to a 1-cochain, for example the line from $g_0$ to $g_1$ represents the 1-cochain $\nu_1(g_0,g_1)$. The brunching structure is such that we label vertices by numbers and draw arrows on links along increasing direction. The graphical representation of the cocycle condition follows from the fact that this 1D complex is the edge of a 2D complex, i.e. the triangle $(g_0,g_1,g_2)$, hence it is closed. Eq.~(\ref{1cocycle}) then states that the (oriented) product of 1-cocycles  along a closed 1D manifold is always trivial. 
 
 Similarly in $d=2$, a 2-cocycle $\nu_{2}(g_1,g_2,g_3)$ satisfies
 \be
 \label{2cocyclee}
 ~\frac{\nu_2(g_1,g_2,g_3)\nu_2(g_0,g_1,g_3)}{\nu_2(g_0,g_2,g_3)\nu_2(g_0,g_1,g_2)}=1=~\raisebox{-.5\height}{\includegraphics[scale=0.7]{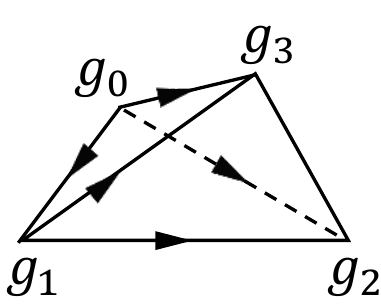}}.
 \ee
 The surface of the 3D tetrahedron $(g_0,g_1,g_2,g_3)$ is a closed 2D manifold, and Eq.~(\ref{2cocyclee}) sates that the oriented product of 2-cocycles on a closed 2D manifold is trivial. In general, the (oriented) product of $d$-cocycles on a closed $d$-dimensional manifold is trivial.
 
 For $d=1$,  1-coboundaries are $\lambda_1(g_0,g_1)=\mu_0(g_1)/\mu_0(g_0)$. A 2-coboundary is 
 \be
 \label{2cob}
 \lambda_2(g_0,g_1,g_2)=\mu_1(g_1,g_2) \mu_1(g_0,g_1)/\mu_1(g_0,g_2).
 \ee 
 The triangle $(g_0,g_1,g_2)$ in Eq.~(\ref{1cocycle}) can be viewed as a 2D membrane associated with a 2-cochain $\nu_2(g_0,g_1,g_2)$. When this 2-cochain is constructed from the 1-chains on the lines on its boundary, then it is a coboundary. If these 1-chains are cocycles, then Eq.~(\ref{1cocycle}) states that this coboundary is trivial, namely the coboundary of a cocycle is trivial, $(d_1 \nu_1)(g_0, g_1,g_2)=1$. A 3-cobondary is 
 \bea
 \label{3cob}
 \lambda_3(g_0,g_1,g_2,g_3)=(d_2 \mu_2)(g_0, g_1,g_2,g_3)\nonumber \\
 =\frac{\mu_2(g_1,g_2,g_3)\mu_2(g_0,g_1,g_3)}{\mu_2(g_0,g_2,g_3)\mu_2(g_0,g_1,g_2)}.
 \eea
 Graphically, the 3D body of the tetrahedron $(g_0,g_1,g_2,g_3)$ in Eq.~(\ref{2cocyclee}) can be associated with a 3-cochain. When constructed from 2-cochains on its 4 faces, we obtain a co-boundary. Eq.~(\ref{2cocycle}) says that the 3-coboundary is trivial if it is constructed out of 2-cocycles, $(d_2 \nu_2)(g_0, g_1,g_2,g_3)=1$.
 
 \subsection{Partition functions and ideal SPT wave function}
 Let us consider a $d+1$-dimensional complex $ M_{\rm{ext}}$, containing $N_v$ sites, as a statistical mechanical model. At each site we have a $g-$valued ``spin". Using $d+1$-cocycles we write a ``partition function" 
 \bea
 \label{eq:Z}
 Z=|G|^{-N_v} \sum_{\{g_i \}} e^{-S(\{ g_i \})},~~~e^{-S(\{ g_i \})}\\
 =\prod_{ik \dots l}\nu^{s_{ij \dots k}}_{1+d}(g_i, g_j  \dots g_k). \nonumber
 \eea
 Here $s_{ij \dots k}=\pm 1$ depends on the orientation of the simplex~\cite{chen2013symmetry}. Due to the cocycle condition this partition function is trivial (=1) if $ M_{\rm{ext}}$ is a closed $d+1$ dimensional manifold. Otherwise, it yields a nontrivial theory on the $d-$dimensional edge of $ M_{\rm{ext}}$, %
 denoted $\partial M_{\rm{ext}}$. The ideal SPT wave function is
 \bea
 &&\Psi(\{g_i \}_{\partial M_{\rm{ext}}})\\
 &&=\mathcal{N}\sum_{g_i \in M_{\rm{ext}} \backslash \partial M_{\rm{ext}} }  \prod_{ik \dots l}\nu^{s_{ij \dots k}}_{1+d}(g_i, g_j  \dots g_k) |\{g_i \} \rangle. \nonumber
 \eea
 Here one sums over the  $N_v^{int}$ internal vertices, in $M_{\rm{ext}}$ not including the boundary $ \partial M_{\rm{ext}}$ where the SPT lives.
 The normalization factor is $\mathcal{N}=|G|^{-N_v^{int}-N_{edge}/2}$. We emphasize that this wave function does not depend on the triangulation and internal structure of $M_{\rm{ext}}$. (In fact it even does not depend on the values of the internal $g's$). For a 1D SPT with $N_{ext}=N$ sites, taking the simplest triangulation of $M_{\rm{ext}}$ such that it contain one internal vertex $g^*$, the SPT wave function can be written as
 \begin{equation} 
 \label{2cocycle}
 \Psi(\{g_i \})=\mathcal{N}\frac{\prod_{i=1}^{N-1} \nu_2(g_i,g_{i+1},g^*)}{\nu_2(g_1,g_N,g^*)}|\{g_i \} \rangle=
 \raisebox{-.5\height}{\includegraphics[scale=0.5]{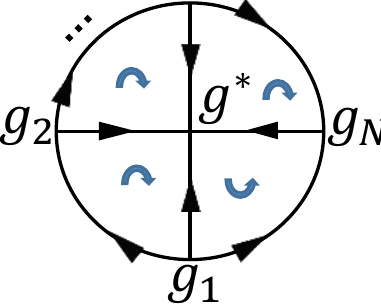}}.
 \end{equation}
 The symmetry has an onsite decomposition $U(g) = \prod_{i \in \partial M_{\rm{ext}}} u_i(g)$, with on-site action $u(g) |g_i \rangle=|g g_i \rangle$. One can change variables and replace the sum over $g^*$ by $g g^*$ and using the cochain condition Eq.~(\ref{cochain}) one can check that $U(g) | \Psi \rangle = | \Psi \rangle$.
 
 \subsection{Partition functions with defects}
 Let $\mathcal{D}$ be a d-dimensional defect in the $d+1$-dimensional manifold $M_{ext}$. The defect has a specific orientation, such that it gives a direction to any 1D trajectory crossing it. This line goes from the ``inside" to the ``outside" of the defect. It acts on the state by applying a symmetry transformation $g$ on sites on one side of the defect. For the identity element $g=e$ the defect is trivial. We define the partition function or wave function in the presence of a defect $Z(g)$ exactly as in Eq.~(\ref{eq:Z}) except that $d+1$-cocycles corresponding to $d+1$ simplexes that are cut by the defect are modified to $\nu_{1+d} (g_i')$ where in the inside $g_i'=g_i$ while in the outside $g_i'=g g_i$. For example, for a 2-cocycle corresponding to a triangle being cut by a line defect we denote
 \begin{equation} 
 \nu_2(g_0,g g_1, g g_2)=\vcenter{\includegraphics[scale=0.7]{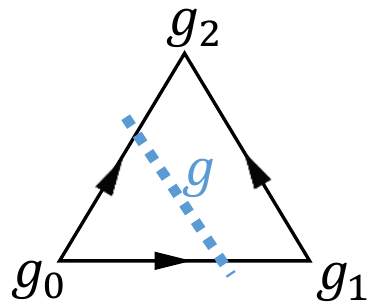}}
 \end{equation} 
 or for a 3-cocycle corresponding to a tetrahedron being cut by a membrane defect,
 \begin{equation} 
 \nu_3(g_0,g_1,g g_2,  g_3)=\vcenter{\includegraphics[scale=0.7]{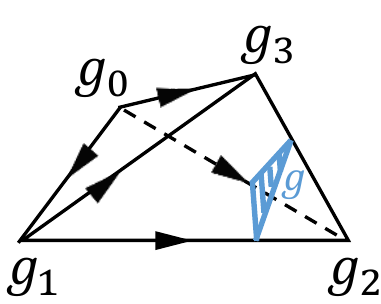}}
 \end{equation} 
 
 A closed defect $\partial \mathcal{D}=0$ is trivial. It transforms sites internal to the manifold by multiplication by $g$. It has no effect since the acttion amplitude does not depend on the internal $g'$s.
 
\section{Equidecomposition Formula for Finite Abelian Groups}
\label{se:prrofcondition}
Let us write a general finite Abelian group $G$ using the group decomposition $G = \intz_{e_{1}} \times \dots \times \intz_{e_l}$, where $e_i$ divides $e_{i+1}$. As shown in  \cite{berkovich1998characters}, the cohomology group has size $|\calh^2[G,U(1)]|=\prod_{i<j} d_{ij}$, where $d_{ij}=gcd(e_i,e_j)$. The cocycles are \cite{huang_wan_ye_2020}
$$
\omega(a,b) = \exp(2 \pi i \sum_{i<j} \frac{p_{ij} a_i b_j}{d_{ij}}),
$$
where $p_{ij}$ enumerates the $\prod_{i<j} d_{ij}$ different cocycles (it is easy to see that apart from the trivial cocycles all the other are non-trivial and form a group. The cocycles conditions can also be verified) and for convenience we set $p_{i\geq j} = 0$.

Using the cocycles, we show that  $f(g) = \sum_s \omega(s,g)\omega^*(g,s) = 0$ for $g \ne e$. This is Eq.~(\ref{DanielEq}), the expression that implies equidecomposition. Substituting the cocycles and writing the sum in an easily calculable form
\begin{gather*}
f(g) = \prod_k \sum_{s_k \in \intz_{e_k}} \left[ \exp\left(2 \pi i \sum_{i} \frac{p_{ki}g_i-p_{ik}g_i}{d_{ik}} \right) \right]^{s_k}
\\
= \prod_k \frac{g(k)^{e_k}-1}{g(k)-1},
\end{gather*}
where $g(k) = \exp\left(2 \pi i \sum_{i} \frac{p_{ki}g_i-p_{ik}g_i}{d_{ik}} \right)$. If $g(k) \neq 1$, it is clear that $f(g)=0$ (as $\frac{e_k}{d_{ik}} \in \intz$ implies $g(k)^{e_k} = 1$). In the other case, it is clear that the sum does not vanish (and is given by $e_k$). The only case that $f(g)=0$ is that if $g(k) \neq 1$ for some $k$. Alternatively, there exists a $k$ such that $ \sum_{i}  \frac{(p_{ki}-p_{ik})g_i}{d_{ik}} \notin \intz$. For example, if all $e_{i} = N$, then $N$ does not divide $\sum_i (p_{ki}-p_{ik})g_i$ for some $k$ is the condition for coboundary invariant $Z_g = 0$. %

\section{Numerical Simulations}
\label{se:numerical}
Here we explain the numerical simulations we have performed in this paper. We describe the numerical procedures used to obtain Table.~\ref{tb:num_zn} and Fig.~\ref{fig3}.

The procedure to obtain Table.~\ref{tb:num_zn} is by direct calculations. As we have seen in the text it is possible to write the 1st \Renyi entropy using the cocycles. 
For each topological phase,  we calculate each $Z_g$ using Eq.~\ref{resiltinter}, and then using Eq.~\ref{charac} we obtain the degeneracies of the 1st moment. We check numerically that these degeneracies are universal by redoing the calculation for  several random coboundaries. %

Plotting Fig.~\ref{fig3} was done by numerically calculating the different types of the effective density matrix. Using Eq.~\ref{eq:SRrhoA}, we calculate the effective density matrix with defect $g$, and by applying Eq.~\ref{charac} we obtain the effective density matrix for the sector $g$. By diagonalizing this effective density matrix, we get the eigenvalues and are able to plot Fig.~\ref{fig3}(a). For example, the random coboundaries used are
\begin{align*}
[0.4114968519411542-0.9114111809949008i,\\ 
-0.39896211626442607-0.9169674093367826i,\\
0.5136056414112213-0.8580263662094353i, \\
0.7505921670806218-0.6607657669077718i]
\end{align*}
for  $\intz_2 \times \intz_2$, where $\beta(g_1,g_2)$ is the line number $1+g_2\cdot N+g_1$ with $N=2$.
For Fig.~\ref{fig3}(b), we need only Eq.~\ref{eq:rhoeff} as we sum all the sectors entanglement. We calculate the entanglement by diagonalizing the effective density matrix, obtaining the eigenvalues, and calculating the entanglement entropy. We do so for many samples, 6100 for $\intz_2 \times \intz_2$ and 2000 for $\intz_3 \times \intz_3$, each with different random coboundary $\beta(g_i)=e^{(2\pi i X_i)}$, where $X_i$ is a random variable drawn from a truncated  normal distribution (between 0 and 1) with zero average and 0.2 standard deviation. Plotting all the samples for the different symmetry groups, we obtain Fig.~\ref{fig3}(b).

\section{Equidecomposition and Maximally Non-Commutative Cocycles Equivalence}
\label{se:eqnmc}
Let us now calculate the size of the group of $g$'s such that $Z_g \neq 0$, which we notate by $G_Z$. $G_Z$ is composed of $\vec{g}$'s that satisfy $B \vec{g} = 0 \pmod {e_l}$ where $B_{ij} = \frac{e_l}{gcd(e_i,e_j)}(p_{ij}-p_{ji})$ is $l\times l$ skew-symmetric matrix. The origin of $B$ is from the negation of Eq.~\ref{condequi}, and we multiplied by $e_l$ to have integer coefficients as it will turn out to be very useful. First, we prove that an $r \times r$ invertible integer matrix $S$ with integer matrix inverse induces isomorphism between $\intz_{m}^r$ to itself by the natural transformation $\vec{h} = S \vec{g} \pmod {m}$. Let $\vec{g_1} \neq \vec{g_2}$ be different vectors with entries in $\intz_m$, then, $\vec{d} \equiv S(\vec{g_1}-\vec{g_2}) \neq 0 \pmod{m}$ as $S$ is invertible, and clearly if $\vec{d}$ is a multiplication of $m$ we have immediately by using $S$ inverse (which is integer matrix by definition) that $\vec{g_1}-\vec{g_2}$ is also a multiplication of $m$ and therefore $0$, contradicting the fact that these are different vectors, therefore, this map is an isomorphism. We continue by using the useful decomposition (over $\intz$), known as Smith normal form, that allows us to decompose $B = TDS$ with matrices over $\intz$ and $T,S$ are invertible with integer matrix inverse while $D$ is diagonal with entries $D_{ii} = \lambda_i$, which are known as invariant factors, that satisfy $\lambda_i$ divides $\lambda_{i+1}$ with possible trailing zeros at the end. We extend $g_i$ to be from $\intz_{e_l}$ by noting that adding $e_i$ will not change that $B\vec{g}=0 \pmod {e_l}$, but we need to compensate the counting with division by $\frac{e_l}{e_i}$ as in Ref.~\cite{huang_wan_ye_2020}. As a result, using our aforementioned isomorphism and counting, we calculate $|G_Z|=\frac{1}{\frac{e_l}{e_1}\dots\frac{e_l}{e_{l-1}}} \# \{ \vec{g} \in \intz_{e_l}^l | D\vec{g}=0 \pmod{e_l} \} = \frac{1}{\frac{e_l}{e_1}\dots\frac{e_l}{e_{l-1}}} \prod_{i,\lambda_i\neq0} \mathrm{gcd}(\lambda_i,e_l) e_l^{l-t}$ where $t$ is the number of non-zero invariant factors of $D$, and the $\mathrm{gcd}$ comes from the well known number of solutions to the linear congruence $ax = 0 \pmod{m}$ which is $\mathrm{gcd}(a,m)$. Combining these results with Ref.~\cite{huang_wan_ye_2020}, we establish a useful relation $|G_0| = |G_Z|$ ($|G_0|$ as defined in Ref.~\cite{huang_wan_ye_2020}, see pages 24-25), and we conclude that equidecomposition, which occurs when $|G_Z|=1$, is equivalent to maximally non-commutative cocycles, which occurs when $|G_0|=1$.


\begin{thebibliography}{48}
\bibitem [{gen()}]{genus}%
	\BibitemOpen
	\href@noop {} {}\bibinfo {note} {The action amplitude may depend on the genus
		of the surface but we will disregard this here.}\BibitemShut {Stop}%
\end{thebibliography}
\end{document}